\documentclass[prfluid, reprint,  amsmath, amssymb,nofootinbib]{revtex4-2}
\usepackage{amsmath, amssymb}
\usepackage[colorlinks,allcolors=blue]{hyperref}
\usepackage{bm,color,xfrac,graphicx}
\usepackage{natbib}
\usepackage[english]{babel}
\graphicspath{{./figures/}}

\DeclareMathAlphabet{\mathsfit}{\encodingdefault}{\sfdefault}{m}{sl}
\SetMathAlphabet{\mathsfit}{bold}{\encodingdefault}{\sfdefault}{bx}{sl}

\newcommand{\tens}[1]{\bm{\mathsfit{#1}}}

\newcommand{\kT}{k_{\mathrm{B}} T}

\newcommand{\ie}{\textit{i.e.}}
\newcommand{\eg}{\textit{e.g.}}
\newcommand{\etc}{\textit{etc.}}
\newcommand{\etal}{\textit{et al.}}

\newcommand{\bigO}{\ensuremath{{\cal O}}}

\newcommand{\grad}[1]{\ensuremath{\boldsymbol{\nabla}{#1}}}

\newcommand{\dive}[1]{\ensuremath{\boldsymbol{\nabla}\cdot{#1}}}
\newcommand{\lapl}[1]{\ensuremath{\nabla^2{#1}}}

\newcommand{\bff}{\ensuremath{\boldsymbol{f}}}
\newcommand{\bbb}{\ensuremath{\boldsymbol{b}}}

\newcommand{\bU}{\ensuremath{\boldsymbol{U}}}

\newcommand{\bI}{\ensuremath{\boldsymbol{I}}}

\newcommand{\be}{\ensuremath{\boldsymbol{e}}}
\newcommand{\bj}{\ensuremath{\boldsymbol{j}}}
\newcommand{\bX}{\ensuremath{\boldsymbol{X}}}

\newcommand{\br}{\ensuremath{\boldsymbol{r}}}
\newcommand{\bn}{\ensuremath{\boldsymbol{n}}}

\newcommand{\bv}{\ensuremath{\boldsymbol{v}}}
\newcommand{\bu}{\ensuremath{\boldsymbol{u}}}

\newcommand{\bd}{\ensuremath{\boldsymbol{d}}}

\newcommand{\bsig}{\ensuremath{\boldsymbol\sigma}}
\newcommand{\bSig}{\ensuremath{\boldsymbol\Sigma}}

\newcommand{\bR}{\ensuremath{\boldsymbol{R}}}

\newcommand{\gamd}{\ensuremath{{\dot{\gamma}}}}
\newcommand{\bzeta}{\ensuremath{\boldsymbol\zeta}}

\newcommand{\tI}{\ensuremath{\tens{I}}}

\newcommand{\tS}{\ensuremath{\tens{S}}}

\newcommand{\Rfu}{\ensuremath{\boldsymbol{R}_{\cal{FU}}}}
\newcommand{\Rfe}{\ensuremath{\boldsymbol{R}_{\cal{F}\mathrm{E}}}}
\newcommand{\Rsu}{\ensuremath{\boldsymbol{R}_{\mathrm{S}\cal{U}}}}
\newcommand{\Rse}{\ensuremath{\boldsymbol{R}_{\mathrm{SE}}}}

\newcommand{\uRfu}{\ensuremath{\boldsymbol{R}^{\mathrm{FU}}}}
\newcommand{\uRfe}{\ensuremath{\boldsymbol{R}^{\mathrm{FE}}}}

\newcommand{\avg}[1]{\ensuremath{\langle{#1}\rangle}}

\newcommand{\pe}{\ensuremath{\mathrm{Pe}}}

\newcommand{\rp}{\ensuremath{\mathrm{p}}}
\newcommand{\rh}{\ensuremath{\mathrm{h}}}
\newcommand{\rf}{\ensuremath{\mathrm{f}}}
\newcommand{\rB}{\ensuremath{\mathrm{B}}}
\newcommand{\rE}{\ensuremath{\mathrm{E}}}
\newcommand{\rP}{\ensuremath{\mathrm{P}}}
\newcommand{\rH}{\ensuremath{\mathrm{H}}}

\begin{document}


\title{Reviving the Suspension Balance Model}

\author{Mu Wang}
\altaffiliation[Current Address: ]{Fabric \& Home Care Innovation Center, The Procter \& Gamble Company, Cincinnati, Ohio 45217, USA}
\affiliation{Divisions of Chemistry and Chemical Engineering and Engineering and Applied Science, California Institute of Technology, Pasadena, California 91125, USA}

\author{Tingtao Zhou}
\email[E-mail: ]{tingtaoz@caltech.edu}
\affiliation
{Divisions of Chemistry and Chemical Engineering and Engineering and Applied Science, California Institute of Technology, Pasadena, California 91125, USA}

\author{John F. Brady}
\email[E-mail: ]{jfbrady@caltech.edu}
\affiliation
{Divisions of Chemistry and Chemical Engineering and Engineering and Applied Science, California Institute of Technology, Pasadena, California 91125, USA}

\date{\today}

\begin{abstract}
The Suspension Balance Model (SBM) [J. Fluid Mech. \textbf{275}, 157 (1994)] for two-phase flows uses the  momentum balance of the particle phase as a closure for the particle flux, showing that particle migration is driven by the divergence of the particle-phase stress. The underlying basis of this model was challenged by Nott~et~al.\ [Phys. Fluids \textbf{23}, 043304 (2011)] where the authors argued that the hydrodynamic contributions to the suspension stress should not appear in the particle-phase momentum balance, being replaced by a different particle-phase stress.  
The particle-phase stress proposed by  Nott~et~al., while mathematically correct, involves the partitioning of the (non-pairwise-additive) hydrodynamic forces, and care is needed to understand how the force on a chosen particle is affected by a second particle.  We show by a simple two-particle calculation what is the proper partitioning, and show that it is consistent thermodynamically and gives the correct equilibrium osmotic pressure of Brownian colloids.   Using Stokesian Dyanmics suspension rheology, we quantitatively demonstrate that the hydrodynamic contribution to the suspension stress is virtually identical to particle-phase stress; the only difference is that the isolated single-particle hydrodynamic stress contribution---the Einstein viscosity correction---must be removed from the suspension stress when used to predict particle flux. Our results validate a key assumption of the SBM and therefore revive its physical foundation.
\end{abstract}

\maketitle

\section{Introduction}
\label{sec:Into}
Concentrated suspensions exhibit rich flow behaviors due to the strong interactions among the dispersed particles.  One example is shear-induced particle migration---the collective particle movement towards regions of low stress or strain rate---in inhomogeneous flows~\citep{shear-struct-susp_acrivos_jor1980}, which is particularly relevant to many applications including micro-filtration~\citep{migration-bidisperse_vdSman_acis2012}, slurry transport~\citep{Koh_experiment-flow-migration_jfm1994}, and biological cell separations~\citep{blood-segregation_graham_softmatt2012}.  Accurate prediction of such migration phenomena from homogeneous rheological measurements is critical for the design and operation of many processes.

The Suspension Balance Model (SBM), first developed by~\citet{NOTjfm94}, quantitatively predicts the shear-induced particle migration in low-Reynolds number ($\mathrm{Re} = a U/\nu_0 \ll 1$) suspensions, where $\nu_0$ is the fluid kinematic viscosity and $a$ and $U$ are the characteristic particle radius and velocity, respectively.  The SBM exploits the mechanical balance of the particle phase as a closure for the particle  flux.  Consequently, the model argues that particle migration is driven by the divergence of the particle-phase stress.  The SBM is equivalent to the phenomenological diffusive-flux model~\citep{shear_migration_leighton_jfm87}, but more robust due to its mechanical origin and the manner in which complex flow geometries can be modeled.  In practice, the SBM agrees well with particle migration measurements in wide-gap Couette devices~\citep{normal-stress-modelling-non-colloidal_morris_jor99, ovarlez2024shear} and pressure-driven flows~\citep{NOTjfm94,Morris_pressure-driven-flow_ijmf1998, orsi2024mass}.    
SBM development has incorporated particle Brownian motion~\citep{migration-brownian-pres_morris_jfm2003,nmr-brownian-migration_codd_pof2009}, size polydispersity~\citep{migration-bidisperse_vdSman_fd2012, howard2022bidisperse}, elasticity~\citep{blood-segregation_graham_softmatt2012}, and flow-arrest transitions~\citep{confine-friction-rheo_garagash_jfm2014}. 

Despite its success in predicting suspension flow behavior, the foundations of the SBM have been called into question, first by Lhuillier~\citep{lhuillier2009migration} and then by~\citet{susbal_nott_physfluid2011}.   A central assumption of the SBM is that the particle-phase stress, $n \avg{\bsig}^\mathrm{p}$, whose divergence drives the shear-induced particle migration, is equal to the particle contribution to the suspension---the mixture of particle plus fluid---stress, $\boldsymbol{\Sigma}^{\mathrm{(p)}}$, which is accessible from rheometry measurements.  The stress $\boldsymbol{\Sigma}^{\mathrm{(p)}}$ is also referred to as the ``particle stress'' by~\citet{stress-system_batchelor_jfm1970}.  Recently, \citet{susbal_nott_physfluid2011} analyzed the suspension mechanics using rigorous spatial averaging~\citep{fluidized-particle_jackson2000}, and claimed that:  (\emph{i}) $n \avg{\bsig}^\mathrm{p} \neq \boldsymbol{\Sigma}^{\mathrm{(p)}}$, and (\emph{ii}) $n \avg{\bsig}^\mathrm{p}$ is not experimentally measurable.  These results call into question the SBM, despite its practical successes, and raise questions about how one is  to model suspension flows more generally.  How are we to predict migration if the quantity needed, the particle-phase stress, $n \avg{\bsig}^\mathrm{p}$,  cannot be measured in an experiment!?


The finding of~\citet{susbal_nott_physfluid2011} that $n \avg{\bsig}^\mathrm{p} \neq \boldsymbol{\Sigma}^{\mathrm{(p)}}$  is due to the fluid-mediated  hydrodynamic interactions (HIs) among the particles (see also \citep{lhuillier2009migration}).  Stress contributions from non-hydrodynamic interactions, such as interparticle colloidal or contact forces, enter both stresses in the same way; if there were no HIs, then the two stresses would be equal. Physically, Nott~\etal's observation is surprising:  why does the presence of a solvent influence in a such fundamental way the particle-phase stress?  

For example, consider the osmotic pressure of  Brownian  suspensions---the force per unit area exerted by the particles on a macroscopic boundary.  The osmotic pressure can be defined and derived from purely thermodynamic arguments---from the variation of the Helmholtz free energy with volume fraction of suspended particles~\citep{soft-matt_doi}.  The incompressible solvent only exerts a constant fluid pressure and has no effect on the osmotic pressure.  Furthermore, the flux of Brownian particles is driven by the gradient in the chemical potential~\cite{brownian-diffusion_batchelor_jfm1976}, which is a thermodynamic variable independent of the dissipative HIs mediated by the fluid, and is directly related to the variation in osmotic pressure with concentration~\citep{soft-matt_doi}.    The viscous solvent can affect the {\em rate} of particle movement, but not the driving force for such movement.  At the same time, \citet{brady1993a} showed that for hard-spheres the Brownian osmotic pressure can be expressed solely in terms of the hydrodynamicly determined Brownian stress, which is expressible in hydrodynamic resistance functions [see Eq.~\eqref{eq:sb}].  According to \citet{susbal_nott_physfluid2011} the hydrodynamic Brownian stress resides in the fluid phase, not the particle phase.  How can these two vastly different interpretations be compatible?


In this work we show that, even though mathematically the exact equality has  not been proven, to an excellent approximation the properly defined particle-phase stress $n \avg{\bsig}^\mathrm{p} = \boldsymbol{\Sigma}^{\mathrm{(p)}} - \boldsymbol{\Sigma}^{\mathrm{(p)}}_0$, where $\boldsymbol{\Sigma}^{\mathrm{(p)}}_0$ is the isolated single-particle hydrodynamic contribution to the suspension stress---the Einstein viscosity correction.   This approximate equality holds for all concentrations regardless of how far from equilibrium the system is driven.  

As we show below, the proper definition comes from a careful examination of how the hydrodynamic force on a particle is affected by a second particle.   There is the direct effect of the force on 1 due to 2, $- \bR_{12}^{\mathrm{FU}}\cdot \bU_2$, where $\bR_{12}^{\mathrm{FU}}$ is the hydrodynamic resistance tensor coupling the force on particle 1 to the velocity of particle 2.  But there is also an indirect effect of how the self-resistance of particle 1 changes due to the presence of particle 2: $- (\bR_{11}^{\mathrm{FU}}- \bR_{11}^{\mathrm{FU}, \infty})\cdot\bU_1$, where $\bR_{11}^{\mathrm{FU},\infty}$ is the isolated particle 1 resistance tensor.  This indirect effect must be included in order to properly assess how particles 1 and 2 interact.  If it is not included, the stress arising from a colliding pair will diverge as the gap between their surfaces goes to zero, whereas the simple physical reasoning shows that the stress should remain finite.  Furthermore, as we show by simulations this partitioning of  the hydrodynamic forces ensures that the particle-phase stress is equal to the equilibrium thermodynamic (or Brownian) stress for colloidal dispersions.  

Thus, in this work we quantitatively validate the central assumption of the SBM and restore it back to a proper model with a sound physical foundation and, more importantly, with experimentally measurable quantities.

The remainder of the paper is arranged as follows.  In Sec.~\ref{sec:balances}, we briefly recall the macroscopic balances for the entire suspension and the particle phase alone, and extend the particle-phase stresses, first defined by \citet{susbal_nott_physfluid2011}, to colloidal Brownian systems.  In Sec.~\ref{sec:part-hydr-inter} we  discuss the partition of the non-pairwise-additive HIs, which is intimately related to the particle-phase stress.  We show at the pair level why the indirect stress contribution must be included, and introduce a new perturbative partition scheme for computing the stress beyond the pair-level.  In Sec.~\ref{sec:method} we describe several pertinent aspects of our computational method.  We present our computational results including equilibrium and non-equilibrium systems in Sec.~\ref{sec:results}.  Finally, we conclude this work with a few closing remarks in Sec.~\ref{sec:conclusions}.

\section{Macroscopic Balances}
\label{sec:balances}

To understand the distinction between the suspension stress and the particle-phase stress, we briefly recall their definitions.  For simplicity, let us consider a suspension of $N$ rigid monodisperse spherical particles of radii $a$ dispersed in a Newtonian fluid of viscosity $\eta_0$, occupying a total volume of $V$.  Since $\mathrm{Re}\ll 1$, the velocity $\bu$ and the pressure $p$ of the fluid are governed by the Stokes equations
\begin{equation}
  \label{eq:stokes}
  \grad{p}= \eta_0 \lapl \bu,\; \dive{\bu} = 0\, ,
\end{equation}
supplemented with no-slip boundary conditions on the particle surfaces.

\subsection{The suspension stress balance}
\label{sec:susp-stress-balance}

Averaging the point-wise equations of motion for the particles and the fluid results in the momentum balance for the suspension---the mixture of particles plus fluid~\citep{stress-system_batchelor_jfm1970}:
\begin{equation}
0 = n \avg{\bbb^\mathrm{ext}} - \grad{} [(1-\phi) \avg{p}^\mathrm{f}] + 2\eta_0 \dive{} \avg{\tens{e}} +  \dive{} \bSig^{(\rp)} \,.
\label{eq:susbal}
\end{equation}
In Eq.~\eqref{eq:susbal}, $\avg{\bbb^\mathrm{ext}}$ is the average net external body force acting on the particles, \eg\ the buoyant force of gravity, $\avg{p}^\rf$ is the average pressure in the fluid,  $\avg{\tens{e}}$ is the average rate of strain in the suspension (the particles plus the fluid), $n = N/V$ is the number density of particles, $\phi= 4n a^3/3$ is the particle volume fraction, and $\bSig^{(\rp)}$ is the particles' contribution to the suspension stress.  We are interested in the stress in at low Reynolds numbers and therefore have set the inertial terms to zero. (This restriction is easily relaxed and does not affect any of the conclusions reached in this work.)  Note that the last term in Eq.~\eqref{eq:susbal} is the leading order term in a gradient expansion of the solid-phase stress tensor, and represents a pointwise uniform stress state in the suspension~\citep{susbal_nott_physfluid2011}.

The mixture is also incompressible 
\begin{equation}
\nabla \cdot \avg{\bu} = 0 \, ,
\label{eq:continuity}
\end{equation}
with the average suspension velocity $\avg{\bu} = \phi \avg{\bu}^\rp + (1-\phi) \avg{\bu}^\rf$.

The particle stress $\bSig^{(\rp)}$ has three distinct contributions,
\begin{equation}
  \label{eq:susp-stress}
  \bSig^{(\rp)} = n\avg{\tens{S}^\rP} + n\avg{\tens{S}^\rH} + n\avg{\tens{S}^\rB}\, ,
\end{equation}
where $n\avg{\tens{S}^\rP}$ arises from interparticle forces such as electrostatic, DLVO and frictional contact forces,  $n\avg{\tens{S}^\rH}$ is the direct flow contributions due to the HIs mediated by the solvent, and $n \avg{\tens{S}^\rB}$, unique to colloidal suspensions, is from the particle Brownian motions.  

The interparticle stress $n\avg{\tens{S}^\rP} = -n\avg{\br \bff^{\rP}}$, where $\bff^\rP$ is the nonhydrodynamic interparticle forces acting on the particles and $\br$ is their positions.  Assuming the interparticle forces are pairwise and sum to zero (net interparticle forces are part of the net ``external'' body force $\avg{\bbb^{\mathrm{ext}}}$), this stress is more typically written as the virial
\begin{equation}
\label{eq:xF}
n\avg{\tens{S}^\rP} = -\frac{1}{2V}\sum_{i=1}^N \sum^N_{\substack{j=1\\ i\neq j}} \br_{ij} \bff^\rP_{ij}\, ,
\end{equation}
where $\br_{ij}=\br_i-\br_j$ is the vector pointing from particle $j$ located at $\br_j$ to particle $i$ located at $\br_i$, and $\bff^\rP_{ij}$ ($=-\bff^\rP_{ji}$) is the nonhydrodynamic interparticle force exerted on particle $i$ due to particle $j$.  

The hydrodynamic flow stress $n\avg{\tens{S}^\rH}$ arises from the rigidity of the suspended particles in the fluid.  For particle $i$ located at $\br_i$, it can be computed by integrating the spatial moment of the surface traction with respect to the particle center over the particle surface $A_i$,
\begin{equation}
\tens{S}_i^\rH =  \int_{A_i} {\tfrac{1}{2}}[(\br-\br_i) \boldsymbol{f} + \boldsymbol{f} (\br-\br_i) ] dS_r \, ,
\label{eq:stresslet}
\end{equation}
where $\br$ is the position on the particle surface and $\boldsymbol{f}=\bsig\!\cdot\!\bn$ is the local surface traction with $\bsig = -p\tI + \eta_0[\grad{\bv} + (\grad{\bv})^\dagger]$ the fluid stress tensor and $\bn$ the unit normal on the particle surface.  Here, $\tI$ is the idem tensor and the symbol~$\dagger$ represents transposition.  The surface traction $\boldsymbol{f}=\boldsymbol{f}(\br; \br^N)$ depends on the local position $\br$ and the configuration of the entire suspension $\br^N=(\br_1, \br_2, \cdots, \br_N)$.  Moreover, integrating $\boldsymbol{f}$ over the surface of particle $i$ leads to the hydrodynamic force $\bff^\mathrm{h}_i$ and the hydrodynamic torque $\boldsymbol{\tau}^\mathrm{h}_i$:
\begin{align}
  \label{eq:hydro-f-t}
\bff^\mathrm{h}_i & = \int_{A_i}\boldsymbol{f} dS_r\, ,\\
\boldsymbol{\tau}^\mathrm{h}_i &=  \int_{A_i}(\br-\br_i)\times\boldsymbol{f} dS_r\, .
\end{align}

From the linearity of the Stokes equations [Eq.~\eqref{eq:stokes}] the hydrodynamic force $\bff^\rh$, torque $\boldsymbol{\tau}^\rh$, and stresslet $\tS^\rH$ on the particles are linearly related to the particle linear and angular velocities and the rate of strain tensor through the grand resistance tensor $\boldsymbol{\mathcal{R}}$,
\begin{equation}
  \label{eq:mob-res-fs}
    \begin{bmatrix}
\boldsymbol{\mathcal{F}}^\rh\\
    \tS^\rH
\end{bmatrix}
=-\boldsymbol{\mathcal{R}}\cdot
\begin{bmatrix}
    \boldsymbol{\mathcal{U}}-\avg{\boldsymbol{\mathcal{U}}}\\
    - \avg{\tens{e}}
\end{bmatrix}.
\end{equation}
Here, the generalized force $\boldsymbol{\mathcal{F}}^\rh = (\bff^{\rh}, \boldsymbol{\tau}^{\rh})^\dagger$ and the generalized velocity $\boldsymbol{\mathcal{U}} = (\bu,\boldsymbol{\omega})^\dagger$ with $\bu$ and $\boldsymbol{\omega}$ respectively the linear and angular velocities of all particles in the suspension.  We adopt a shorthand notation for $\bff^{\rh}$, $\boldsymbol{\tau}^{\rh}$, $\bu$, and $\boldsymbol{\omega}$ such that, for example, $\bff^{\rh}=(\bff^\rh_1, \bff^\rh_2, \cdots, \bff^\rh_N)^\dagger$.  The grand resistance tensor $\boldsymbol{\mathcal{R}}$ depends only on the suspension configurations, and, as is customary, can be partitioned as
\begin{equation}
  \label{eq:res-part}
\boldsymbol{\mathcal{R}} = 
  \begin{bmatrix}
    \Rfu & \Rfe \\
    \Rsu & \Rse
  \end{bmatrix},
\end{equation}
where, for example, $\Rfu$ is the resistance functions describing the coupling between the generalized force $\boldsymbol{\mathcal{F}}^\rh$ and the generalized velocity disturbance $\boldsymbol{\mathcal{U}}-\avg{\boldsymbol{\mathcal{U}}}$.  Because HIs are long-range and non-pairwise-additive, each element of $\boldsymbol{\mathcal{R}}$ depends on the configuration of all particles $\br^N$.  With the resistance functions, the hydrodynamic stress $n\avg{\tS^\rH}$, without external forces and torques, can be expressed as
\begin{equation}
n\avg{\tS^\rE} = - n\avg{\Rsu\cdot\Rfu^{-1}\cdot\Rfe-\Rse}:\avg{\tens{e}}\, .
\label{eq:stresslet}
\end{equation}
We have  written the hydrodynamic stresslet $n\avg{\tS^\rH} = n\avg{\tS^\rE}$ to reflect the fact that this term is directly proportional to the suspension rate of strain $\avg{\be}$.  In going forward the superscript $\rE$ in place of $\rH$ denotes hydrodynamic stresses that are directly proportional to the macroscopic strain rate $\avg{\tens{e}}$.

The Brownian stress $n\avg{\tens{S}^\rB}$ is significant when the particles are sufficiently small ($\le 1\mu\mathrm{m}$).  The solvent thermal fluctuations  exert a Brownian force $\boldsymbol{\mathcal{F}}^\rB$ on the particles satisfying the fluctuation-dissipation theorem
\begin{equation}
  \label{eq:fb}
 \overline{\boldsymbol{\mathcal{F}}^\rB(t)}=0 \text{ and }\overline{\boldsymbol{\mathcal{F}}^\rB(0)\boldsymbol{\mathcal{F}}^\rB(t)}=2\kT\Rfu\delta(t)\, ,
\end{equation}
where the overline indicates time averaging and $\delta(t)$ is the Dirac delta function.  Although the average of the Brownian force is zero, it results in a non-zero Brownian stress contribution.  Averaging the stress from the Brownian force $\boldsymbol{\mathcal{F}}^\rB$ over a time short compared to the particle configuration change but long relative to the thermal fluctuations, \citet{bossisbrady89} showed that
\begin{equation}
  \label{eq:sb}
n\avg{\tens{S}^\rB} = -n\kT\tI - n\kT \langle \grad{}_c \cdot (\Rsu\cdot\Rfu^{-1}) \rangle\, ,
\end{equation}
where $\grad{}_c\cdot$ is the configurational divergence of $\br^N$ acting on the last index of the tensor.  The first term on the right side of Eq.~\eqref{eq:sb} is the kinetic contribution from the particle Brownian motion, while the second term arises from the HIs.

The non-kinetic Brownian stress arises from  hydrodynamic flows generated by the Brownian forces.  From Eq.~(\ref{eq:mob-res-fs}) $n\langle\tS^\rH\rangle = - n\langle \Rsu\cdot\Rfu^{-1}\cdot \boldsymbol{\mathcal{F}}^\rB\rangle$, which when averaged over the thermal fluctuations gives  the second term on the RHS of Eq.~\eqref{eq:sb}.  In a similar manner, there are hydrodynamic flows and stresses associated with the interparticle forces $n\langle\tS^\rH\rangle = - n\langle \Rsu\cdot\Rfu^{-1}\cdot \bff^P\rangle$, which could be added to $\tens{S}^\rP$. Such an addition was used in our earlier work.  Here, however, in order to be consistent with the notation of \citet{susbal_nott_physfluid2011} only the direct interparticle force contribution, the virial, is included in $\tens{S}^\rP$.

\subsection{The particle-phase stress balance}
\label{sec:part-phase-stress}

The macroscopic momentum balance on the particles alone---the particle-phase balance---is~\citep{susbal_nott_physfluid2011}:
\begin{equation}
0 = n\avg{\bff^\rh}_\mathrm{drag} + n\avg{\bbb^{\mathrm{ext}}} + \nabla\cdot (n\avg{\bsig}^{\rp}) \, ,
\label{eq:parbal}
\end{equation}
where $n\avg{\bff^\rh}_\mathrm{drag}$ is the average hydrodynamic drag force, 
and $n\avg{\bsig}^{\rp}$ is the particle phase stress.  Eq.~\eqref{eq:parbal} arises from a rigorous averaging process on the particle  equations of motion and is the main result of~\citet{susbal_nott_physfluid2011}.  

The hydrodynamic drag force is proportional (to leading order in spatial gradients) to the difference between the average particle velocity $\avg{\bu}^\rp$ and that of the suspension $\avg{\bu}$:
\begin{equation}
n\avg{\bff^\rh}_\mathrm{drag} = - n \bzeta \cdot( \avg{\bu}^\rp - \avg{\bu}) \, ,
\label{eq:dragflux}
\end{equation}
where $\bzeta(\phi)$ is the average drag (tensor) on the particle phase, which can be measured, for example, in a sedimentation experiment and calculated in a simulation as discussed below.   Equation (\ref{eq:parbal}) then serves to determine the relative particle flux
\begin{equation}
 \bj_{rel} \equiv n (\avg{\bu}^\rp - \avg{\bu})  = \bzeta^{-1} \cdot [n\avg{\bbb^{\mathrm{ext}}} + \nabla\cdot (n\avg{\bsig}^{\rp})] \, ,
 \label{eq:jrel}
 \end{equation}
 which is to be used in the conservation equation for particle number density
\begin{equation}
\frac{\partial n}{\partial t} + \avg{\bu} \cdot\nabla n = - \nabla \cdot \bj_{rel} \, .
\label{eq:numberconservation}
\end{equation}
This is the essence of the SBM---it provides a constitutive equation for the particle flux in terms of measurable quantities:  (\textit{i}) drag  $\bzeta$, (\textit{ii}) body force $n\avg{\bbb^{\mathrm{ext}}} $  and (\textit{iii}) particle-phase stress $n\avg{\bsig}^{\rp}$.  The evolution of the number density  $n$ is then used along with the suspension momentum balance Eq.~(\ref{eq:susbal}) to determine flows and forces on boundaries.  (Boundary conditions for the balance equations are also needed, but this is a topic for a different discussion.)

The external body force needs no further elaboration.  The drag force (\ref{eq:dragflux}) is simply the average hydrodynamic force
\begin{equation}
n\avg{\bff^\rh}_\mathrm{drag}  = \frac{1}{V} \sum_i \bff_i^h \, ,
\label{eq:dragforce}
\end{equation}
where, from the grand resistance tensor (\ref{eq:mob-res-fs}),  the hydrodynamic force on particle $i$ is
\begin{equation}
  \label{eq:HIforceoni}
  \begin{split}
  \bff^\mathrm{h}_i = & -  \sum_{j \ne i} [\uRfu_{ij} \cdot (\bu_j - \langle \bu \rangle)
  \\  & \quad + \bR^{\mathrm{F\Omega}}_{ij} \cdot (\boldsymbol{\omega}_j - \langle \boldsymbol{\omega} \rangle)  
  - \uRfe_{ij} : \langle \tens{e} \rangle] \, ,\\
  \end{split}
\end{equation}
where the hydrodynamic resistance functions $\uRfu_{ij}$, $\bR^{\mathrm{F\Omega}}_{ij}$, and $\uRfe_{ij}$, respectively, describe the coupling between the force on particle~$i$ due to the linear velocity, angular velocity, and the strain rate of particle~$j$. For the grand resistance tensor $\boldsymbol{\mathcal{R}}$ in Eq.~\eqref{eq:res-part}, $\uRfu_{ij}$ and $\bR^{\mathrm{F\Omega}}_{ij}$ are submatrices of $\Rfu$, and $\uRfe_{ij}$ is a submatrix of $\Rfe$.  Each resistance function in Eq.~\eqref{eq:fp-nott} depends on $\br^N$, the configuration of the entire suspension.

In Stokes flow, it is possible to define a `center of resistance' analogous to the center of mass, and define motion relative to it 
\begin{equation}
\bu_j = \bu_{\mathrm{CR}} + \bu^\prime_j \ ; \ \boldsymbol{\omega}_j = \boldsymbol{\omega}_{\mathrm{CR}} + \boldsymbol{\omega^\prime}_j\ ,
\label{eq:CR}
\end{equation}
such that 
\begin{equation}
\sum_{ij} \uRfu_{ij} \cdot \bu^\prime_j = 0 \ , \ \sum_{ij} \bR^{\mathrm{F\Omega}}_{ij} \cdot \boldsymbol{\omega^\prime}_j = 0\ .
\label{eq:CR1}
\end{equation}
Thus, one identifies $\avg{\bu}^\rp = \bu_{\mathrm{CR}}$ and the average drag tensor is $\bzeta =\sum_{ij} \uRfu_{ij}$~\footnote{The drag coefficient $\bzeta$ as defined is for a ``resistivity problem,'' e.g. for flow though a fixed bed or porous medium---the particles are fixed in space and the fluid flows past. ``Freely mobile'' suspensions correspond to a mobility problem in Stokes flow.  The two drag coefficients are not the same, but do not differ significantly except for  dilute systems.} .  Similarly,  $\bzeta_\mathrm{TR} =\sum_{ij} \bR^{\mathrm{F\Omega}}_{ij} $ and $\bzeta_\mathrm{TE} =  \sum_{ij} \uRfe_{ij}$, the drag coefficients coupling angular velocity to translational flux, etc.  Note that $\bzeta_\mathrm{TR}$ must be a chiral second order tensor and therefore the suspension microstructure must be chiral in order for there to be a coupling between translation and rotation.  The microstructure must also support a third order tensor in order for $\bzeta_\mathrm{TE}$ to be nonzero.  In (\ref{eq:dragflux}) we have just shown the contribution from $\bzeta$, which is present in any microstructure.  


For colloidal suspensions the particle-phase stress  has three distinct contributions:
\begin{equation}
  \label{eq:parstress}
n\avg{\bsig}^{\rp} = n\avg{\bsig^\rP}^{\rp}+n\avg{\bsig^\rH}^{\rp}+n\avg{\bsig^\rB}^{\rp}.
\end{equation}
The right hand side of Eq.~\eqref{eq:parstress} represents contributions from the contact or the interparticle forces, the direct HIs, and the particle Brownian motions, respectively.  \citet{susbal_nott_physfluid2011} examined the non-Brownian stress contributions $n\avg{\bsig^\rP}^{\rp}$ and $n\avg{\bsig^\rH}^{\rp}$, but did not carefully consider the Brownian particle-phase stress $n\avg{\bsig^\rB}^{\rp}$.

\citet{susbal_nott_physfluid2011} showed that the suspension and the particle-phase stress due to the interparticle or contact forces are the same, \ie, $n\avg{\bsig^\rP}^{\rp}=n\avg{\tS^\rP}$.  As expected, this stress contribution arises entirely from the particles and is not affected by the interstitial fluid.

The direct hydrodynamic contribution to the particle phase stress, $n\avg{\bsig^\rH}^{\rp}$, is 
\begin{equation}
\label{eq:p-stress}
n\avg{\bsig^{\rH}}^{\mathrm{p}}  = 
-\frac{1}{2V}\sum_{i=1}^N\sum_{\substack{j=1\\ i\neq j}}^N \br_{ij} \bff_{ij}^{\mathrm{h}}\, ,
\end{equation}
which is the spatial moment of the ``pairwise'' hydrodynamic forces $\bff^\rh_{ij}$, and is similar to the stress contributions from the interparticle or the frictional contact forces.  However, the ``pairwise'' hydrodynamic force on particle $i$ due to particle $j$, $\bff^\mathrm{h}_{ij}$, is not uniquely defined since the HIs are dissipative and non-pairwise-additive.  On the other hand, regardless of the pairwise partition, $\bff^\mathrm{h}_{ij}$ must satisfy 
\begin{equation}
  \label{eq:p-hydro-f}
\sum_{j=1}^N \bff^\rh_{ij}=\bff^\mathrm{h}_{i}\, ,  
\end{equation}
where $\bff^\mathrm{h}_{i}$ is the total hydrodynamic force on particle $i$ and is given by (\ref{eq:HIforceoni}).   Note that the summation of Eq.~\eqref{eq:p-hydro-f} is unrestricted and therefore also includes the self contribution $\bff^\rh_{ii}$. 

Following  from (\ref{eq:HIforceoni}) \citet{susbal_nott_physfluid2011}  proposed the  pairwise partition of the HIs:
\begin{equation}
  \label{eq:fp-nott}
  \bff^\mathrm{h}_{ij}= \uRfu_{ij} \cdot (\bu_j - \langle \bu \rangle)
  + \bR^{\mathrm{F\Omega}}_{ij} \cdot (\boldsymbol{\omega}_j - \langle \boldsymbol{\omega} \rangle)  
  - \uRfe_{ij} : \langle \tens{e} \rangle\, .
\end{equation}
Eq.~\eqref{eq:p-stress} and \eqref{eq:fp-nott} show that the particle phase hydrodynamic stress $n\avg{\bsig^\mathrm{H}}^\mathrm{p}$ is linear in the particle kinematics, which  allows us to introduce new resistance functions $\boldsymbol{\mathfrak{R}}_{\sigma{\cal U}}$ and $\boldsymbol{\mathfrak{R}}_{\sigma\rE}$ such that,
\begin{equation}
  \label{eq:stresslet-E}
  n\avg{\bsig^\mathrm{E}}^\mathrm{p} = -n \langle \boldsymbol{\mathfrak{R}}_{\sigma{\cal U}} \cdot \Rfu^{-1}\cdot \Rfe -\boldsymbol{\mathfrak{R}}_{\sigma\mathrm{E}} \rangle :\avg{\tens{e}}\, ,
\end{equation}
for suspensions under linear flow without external forces and torques.  For the particle pair  $(i,j)$ the submatrices of $\boldsymbol{\mathfrak{R}}_{\sigma{\cal U}}$ contain 
\begin{equation}
  \label{eq:rsigu}
\boldsymbol{\mathfrak{R}}^{\sigma\mathrm{U}}_{ij}=\tfrac{1}{2}\br_{ij} \uRfu_{ij} \text{ and }
\boldsymbol{\mathfrak{R}}^{\sigma\Omega}_{ij}=\tfrac{1}{2}\br_{ij}  \bR^{\mathrm{F\Omega}}_{ij},
\end{equation}
and the submatrix of $\boldsymbol{\mathfrak{R}}_{\sigma\rE}$ is
\begin{equation}
  \label{eq:rsige}
  \boldsymbol{\mathfrak{R}}^{\sigma\rE}_{ij}=\tfrac{1}{2}\br_{ij} \uRfe_{ij}\, .
\end{equation}
The definition of $n\avg{\bsig^\rH}^\rp$ in Eq.~\eqref{eq:p-stress} (or $n\avg{\bsig^\rE}^\rp$ in Eq.~(\ref{eq:stresslet-E})) suggests that there is no self contribution to the particle phase stress, \ie, $\boldsymbol{\mathfrak{R}}^{\sigma\mathrm{U}}_{ii}=0$, $\boldsymbol{\mathfrak{R}}^{\sigma\Omega}_{ii}=0 $, and $\boldsymbol{\mathfrak{R}}^{\sigma\rE}_{ii}=0$, which follows because $\br_{ii} = 0$.

Although very similar in structure, the direct hydrodynamic stress in the suspension $n\avg{\tS^\rH}$, Eq.~(\ref{eq:stresslet}),  and in the particle phase $n\avg{\bsig^\rH}^\rp$, Eq.~(\ref{eq:stresslet-E}),  are different.  Since the particle migration is driven by the stress in the particle phase, the difference prompted \citet{susbal_nott_physfluid2011} to realize that the original SBM incorrectly assumes that the two stresses are the same.

In colloidal suspensions, the particle Brownian motion also contributes to the particle-phase stress as $n\avg{\bsig^\rB}^\rp$.  This contribution is evaluated using the same time averaging procedure that lead to Eq.~\eqref{eq:sb}, but with the resistance function $\boldsymbol{\mathfrak{R}}_{\sigma{\cal U}}$.  Following \citet{bossisbrady89}, this is
\begin{equation}
  \label{eq:sigB}
  n\avg{\bsig^\mathrm{B}}^\mathrm{p} = -n\kT\tI -n \kT \langle \grad{}_c \cdot ( \boldsymbol{\mathfrak{R}}_{\sigma{\cal U}} \cdot\Rfu^{-1}) \rangle\, .
\end{equation}
On the right hand side of Eq.~\eqref{eq:sigB}, the first term is due to particle velocity fluctuations, and the second term arises from the fluctuating suspension configurations.  \citet{susbal_nott_physfluid2011} briefly considered the Brownian contribution to the particle phase stress, but  concluded that the particle phase only contributes $-n\kT\tI$ to the suspension Brownian stress.  They overlooked the contributions from the fluctuating configurations in the particle phase.

\section{Partitioning the hydrodynamic forces}
\label{sec:part-hydr-inter}

Sec.~\ref{sec:part-phase-stress} demonstrates that partitioning the hydrodynamic forces is crucial for the direct hydrodynamic and the Brownian particle-phase stresses $n\avg{\bsig^\rH}^\rp$  and $n\avg{\bsig^\rB}^\rp$.  
According to Eq.~\eqref{eq:p-stress}, what is needed is the hydrodynamic force on particle `$i$' due to particle `$j$'.  Let us consider only two particles 1 and 2 and just focus on the $\mathrm{FU}$ coupling.  Following Nott~\etal\nocite{susbal_nott_physfluid2011}  in  Eq.~\eqref{eq:fp-nott} for only two particles the contribution to the hydrodynamic particle stress has the form
\begin{eqnarray}
\avg{\bsig^{\rH}}^{\mathrm{p}} & : \Rightarrow & {\textstyle{\frac{1}{2}}} \br_{12}(\bR_{12}^{\mathrm{FU}}\!\cdot\!\bU_2 - \bR_{21}^{\mathrm{FU}}\cdot\!\bU_1) \nonumber \\*
&  & =  {\textstyle{\frac{1}{2}}} \br_{12}\bR_{12}^{\mathrm{FU}}\!\cdot\!(\bU_2 - \bU_1)\, ,
\label{eq:hydrostresspair}
\end{eqnarray}
where we have used the symmetry $\bR_{\alpha \beta}^{\mathrm{FU}}  = \bR_{3- \alpha 3- \beta}^{\mathrm{FU}}$ of the resistance matrix.  To this stress we add the virial stress arising from nonhydodynamic interparticles forces,  Eq.~\eqref{eq:xF},
\begin{equation}
\avg{\bsig^{\rP}}^{\mathrm{p}}  : \Rightarrow  - {\textstyle{\frac{1}{2}}} \br_{12}\ 2\bff^P_{12} \, ,
\label{eq:nonhydrostresspair}
\end{equation}
to give 
\begin{equation}
\avg{\bsig^{\rH}}^{\mathrm{p}} + \avg{\bsig^{\rP}}^{\mathrm{p}}  : \Rightarrow  {\textstyle{\frac{1}{2}}} \br_{12}[\bR_{12}^{\mathrm{FU}}\!\cdot\!(\bU_2 - \bU_1) - 2\bff^P_{12}]\, .
\label{eq:Nottstress}
\end{equation}
From the equations of motion for the two particles under the action of the interparticle force, we can solve for $2\bff^P_{12}$ in terms of the relative velocity so that Eq.~\eqref{eq:Nottstress} becomes
\begin{equation}
\avg{\bsig^{\rH}}^{\mathrm{p}} + \avg{\bsig^{\rP}}^{\mathrm{p}}  : \Rightarrow  {\textstyle{\frac{1}{2}}} \br_{12}\bR_{22}^{\mathrm{FU}}\!\cdot\!(\bU_2 - \bU_1) \, .
\label{eq:Nottstress2}
\end{equation}

Considering motion along the line of centers  Eq.~\eqref{eq:Nottstress2} becomes
\begin{equation}
\avg{\bsig^{\rH}}^{\mathrm{p}} + \avg{\bsig^{\rP}}^{\mathrm{p}} : \Rightarrow - {\textstyle{\frac{1}{2}}} r_{12}X_{22}^A U_{rel} \bd\bd\, ,
\label{eq:hydrostresspair2}
\end{equation}
where $U_{rel}= (\bU_2 - \bU_1)\cdot\bd$ is the relative velocity along the line of centers characterized by the unit vector $\bd = (\br_2 - \br_1)/r_{12}$, and $X_{22}^A(r_{12})$ is the scalar hydrodynamic resistance function. 

As the two particles come close to one anther, lubrication interactions dominate and $X_{22}^A \sim  \frac{3}{2}\pi \eta_0 /\epsilon$, as $\epsilon \rightarrow 0$, where $\epsilon = 2(r_{12} - a_1 - a_2)/(a_1 + a_2)$.  Thus the particle-phase  pressure from Eq.~\eqref{eq:hydrostresspair2} scales as
\begin{equation}
 \Pi^{H,\mathrm{p}} \sim  {\textstyle{\frac{3}{4}}} \pi \eta_0 r_{12}U_{rel} /\epsilon\, ,
 \label{eq:hydropressurepair}
\end{equation}
and diverges as the two particles come into contact.  

At the same time, the suspension stress in Eq.~\eqref{eq:susp-stress},  $\bSig^{(\rp)} = n\avg{\tens{S}^\rH} + n\avg{\tens{S}^\rP}$,  arising from both the hydrodynamic stresslet and virial stress, is finite as two particles come into contact:
\begin{eqnarray}
\avg{\tens{S}^\rH} + \avg{\tens{S}^\rP} &: \Rightarrow &  - [(\bR_{22}^{\mathrm{SU}} + \bR_{12}^{\mathrm{SU}} \nonumber \\* & & \quad - {\textstyle{\frac{1}{2}}} \br_{21}(\bR_{12}^{\mathrm{FU}} - \bR_{22}^{\mathrm{FU}})]\cdot (\bU_2 - \bU_1) .\quad \quad
\label{eq:pairsusstress}
\end{eqnarray}
And from the known asymptotic forms for $\bR_{22}^{\mathrm{SU}}$, etc. Eq.~\eqref{eq:pairsusstress} is $O(1)$ as $\epsilon \rightarrow 0$.

 As shown by Nott~\etal\cite{susbal_nott_physfluid2011}, since the suspension stress is equal to the sum of the particle-phase  stress plus the fluid-phase  stress, this means the fluid-phase stress must also diverge as two particles come into contact.  As two particles collide generating a large positive pressure, a large negative pressure is generated in the fluid to suck the fluid into the small gap separating the two particles.  

How would a third particle far removed from this colliding pair respond?  Particle flux is driven by particle-phase stress gradients (c.f. Eq.~\eqref{eq:jrel}) and so the large positive particle-phase pressure will drive the third particle away.  But at the same time the large negative fluid pressure will draw it in.  But the fluid-phase stress is not in the particle conservation equation and therefore cannot counteract the large particle-phase stress. Simulations show that the third particle drifts in the weak collisional-induced force-dipole or stresslet flow that decays as $1/r^2$, just as would be predicted if the suspension, not the particle-phase, stress were responsible for driving particle flux.  

If we look more closely at the hydrodynamic force exerted on particle 1 due to particle 2, we see that there is the `direct' interaction term $- \bR_{12}^{\mathrm{FU}} \cdot\bU_2$ that was used in the particle-phase stress Eq.~\eqref{eq:hydrostresspair}.   But there is also an `indirect' contribution from the interaction between the two particles, $-(\bR_{11}^{\mathrm{FU}} - \bR_{11}^{\mathrm{FU}, \infty})\cdot \bU_1$, measuring how the force on particle 1 due to its own motion changes because of the presence of particle 2.  What we need in the expression for the particle-phase stress Eq.~\eqref{eq:p-stress} is any and all contributions to the hydrodynamic force on particle 1 that are `due' to particle 2.  Clearly, this indirect contribution arises from the presence of particle 2 and is zero if there are no HIs.  

Including this contribution to the force on 1 due to 2 and repeating the steps that led to  Eq.~\eqref{eq:Nottstress2} we now find
\begin{equation}
\avg{\bsig^{\rH}}^{\mathrm{p}} + \avg{\bsig^{\rP}}^{\mathrm{p}}  : \Rightarrow  {\textstyle{\frac{1}{2}}} \br_{12}\bR_{22}^{\mathrm{FU}, \infty}\!\cdot\!(\bU_2 - \bU_1) \, .
\label{eq:notNottstress2}
\end{equation}
There is no longer a divergence in the particle-phase stress as two particle collide.  Rather, the stress is finite just as it is for the suspension stress.  Note, that one recovers the result for the stress  in the absence of HI as $\bR_{22}^{\mathrm{FU}, \infty} = 6 \pi \eta_0 \bI$.

From this two-particle example we argue that one needs to include both the direct and indirect contributions to the force on particle 1 due to particle 2 to get the proper expression for the particle-phase stress.  But how does one do this beyond the two particle level?

Here we show how to partition the HIs based on the idea of configuration perturbation.  This is reminiscent of the test particle insertion technique for computing the chemical potential in Monte Carlo computations~\citep{CompSimLiq}.  The perturbative partition of HIs works as follows.  The hydrodynamic force of particle $i$ in an $N$-particle suspension, $\bff_{i}^\rh=\bff_{i}^\rh(\br^N; \bu, \boldsymbol{\omega},\ldots)$, in general, depends non-linearly  on the suspension configuration $\br^N$, but  depends linearly  on the particle kinematics $\bu$, $\boldsymbol{\omega}$, \etc\  The configuration perturbation can be conveniently computed from the resistance formalism in Eq.~\eqref{eq:mob-res-fs} by removing a particle, say, particle $j$, from the $N$-particle system while keeping the positions and  kinematics of the  remaining particles unchanged.  The HIs of the perturbed system depend on the configuration and kinematics of the remaining particles, $\br^{N-1}$, $\bu'$, and $\boldsymbol{\omega}'$, where $\br^N = \br^{N-1}\cup\br_j$, $\bu=\bu'\cup\bu_j$, $\boldsymbol{\omega}=\boldsymbol{\omega}'\cup\boldsymbol{\omega}_j$, \etc\  The perturbed hydrodynamic force on particle $i$ is $\bff_{i}^\mathrm{h}{}'(j) = \bff_{i}^\mathrm{h}{}'(j)(\br^{N-1};  \bu', \boldsymbol{\omega}')$, and can be computed from the perturbed grand resistance tensor $\boldsymbol{\mathcal{R}}'$.  Note that the resistance functions between particles $p$ and $q$ in the perturbed configuration differ from the unperturbed one, \eg, $ \uRfu_{pq}(\br^N)\neq \uRfu_{pq}(\br^{N-1})$, due to the non-pairwise-additive and long-range nature of HIs.  The difference between $\bff_i^\mathrm{h}$ and $\bff_i^\mathrm{h}{}'(j)$ leads to the perturbative pairwise partition
\begin{equation}
   \label{eq:fij-new}
   \widetilde{\bff}_{ij}^\mathrm{h}=\bff_i^\mathrm{h}-\bff_i^\mathrm{h}{}'(j)\, ;
\end{equation}
that is, the ``pairwise'' force on $i$ due to $j$ is the force on $i$ with and without $j$ present.  For only two particles, this gives directly 
\begin{equation}
   \label{eq:fij-new2}
   \widetilde{\bff}_{12}^\mathrm{h}= - \bR_{12}^{\mathrm{FU}} \cdot \bU_2 - (\bR_{11}^{\mathrm{FU}}  - \bR_{11}^{\mathrm{FU}, \infty})\cdot  \bU_1 \, ,
\end{equation}
which is precisely what we argued gives the correct particle-phase stress at the pair level.

The corresponding particle-phase direct + indirect hydrodynamic stress is thus
\begin{equation}
   \label{eq:sig-new}
n\avg{\widetilde{\bsig}^{\rH}}^{\mathrm{p}}  = 
-\frac{1}{2V}\sum_{i=1}^N\sum_{\substack{j=1\\ i\neq j}}^N \br_{ij} \widetilde{\bff}_{ij}^{\mathrm{h}}\, .
\end{equation}

\begin{figure*}
  \centering
  \includegraphics[width=5.5in]{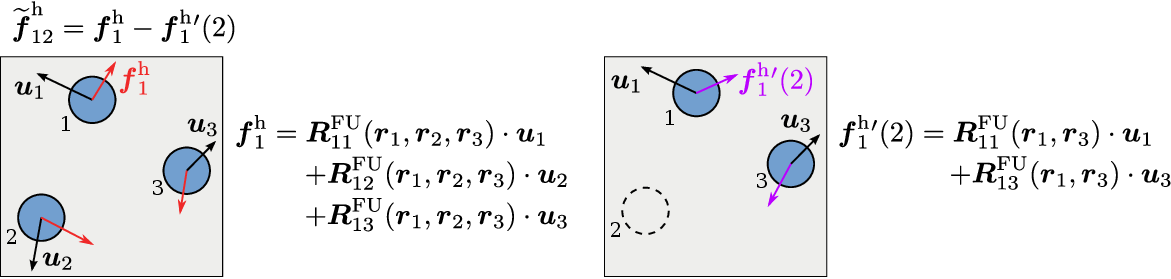}
  \caption{Schematic illustration of computing the pairwise hydrodynamic force $\widetilde{\bff}_{12}^\mathrm{h}$ for a $3$-particle system in a quiescent fluid ($\avg{\bu}=0$).  On the left, the hydrodynamic force on particle $1$, $\bff_1^\rh$, is calculated with all the particles present. On the right, $\bff_1^\mathrm{h}{}'(2)$ is computed with particle $2$ removed (dashed line).  In both cases, the particle velocities are unchanged.
}
  \label{fig:scheme}
\end{figure*}

Fig.~\ref{fig:scheme} illustrates the computation of the perturbative pairwise force $\widetilde{\bff}_{12}^\rh$ in a $3$-particle system with $\avg{\bu}=0$.  The configuration dependence of the resistance functions are explicitly shown.   To evaluate $\widetilde{\bff}_{12}^\rh$, we first compute $\bff^\rh_1$, which depends on the positions and velocities of all three particles.  Next, we remove particle $2$ from the system, and compute the force on particle $1$, $\bff^\rh_1{}'(2)$, which is only a function of the positions and velocities of particles $1$ and $3$.  Finally, the perturbative pairwise force is calculated from their difference $\widetilde{\bff}_{12}^\rh=\bff^\rh_1(\br_1,\br_2,\br_3; \bu_1,\bu_2,\bu_3) - \bff^\rh_1{}'(2)(\br_1,\br_3; \bu_1,\bu_3)$.  Compared to the resistance pairwise force, $\bff_{12}^\rh=\uRfu_{12}(\br_1, \br_2, \br_3)\cdot\bu_2$,  the perturbative pairwise force, $\widetilde{\bff}_{12}^\mathrm{h}$, also includes $[\uRfu_{11}(\br_1, \br_2, \br_3) - \uRfu_{11}(\br_1, \br_3)]\cdot\bu_1$ and $[\uRfu_{13}(\br_1, \br_2, \br_3) - \uRfu_{13}(\br_1, \br_3)]\cdot\bu_3$ resulting from  the removal of particle $2$.

A few comments are in order for the perturbative computation of $\widetilde{\bff}_{ij}^\rh$ in Eq.~\eqref{eq:fij-new}.  First, it is not limited to HIs.  The method is completely general and reduces to the correct result for strictly pairwise interactions.  Second, although there is a self contribution $\widetilde{\bff}_{ii}^\rh$, and it changes with each particle $j$ removal, it is not needed because it is not used in the computation of the particle-phase stress and the total force on particle $i$ is already found from (\ref{eq:HIforceoni}). 
Third, evaluating $\widetilde{\bff}_{ij}^\mathrm{h}$ is computationally expensive, and calculating the perturbative hydrodynamic particle stress $\avg{\widetilde{\bsig}^{\rH}}^{\rp}$ in an $N$-particle system requires HIs of all $N$ perturbed configurations.  On the other hand, $\widetilde{\bff}_{ij}^\rh$ is linear to the particle kinematics such as $\bu$ and $\boldsymbol{\omega}$, and from Eq.~\eqref{eq:sig-new}, new resistance functions $\widetilde{\boldsymbol{\mathfrak{R}}}_{\sigma{\cal U}}$ and $\widetilde{\boldsymbol{\mathfrak{R}}}_{\sigma\mathrm{E}}$ can be introduced from the spatial moment of $\widetilde{\bff}^\rh_{ij}$.  For suspensions in linear flow, the direct + indirect hydrodynamic particle phase stress from the perturbative partition is
\begin{equation}
  \label{eq:newsigE}
  n\avg{\widetilde{\bsig}^\rE}^\rp = -n \langle \widetilde{\boldsymbol{\mathfrak{R}}}_{\sigma{\cal U}} \cdot \Rfu^{-1}\cdot \Rfe -\widetilde{\boldsymbol{\mathfrak{R}}}_{\sigma\mathrm{E}} \rangle :\avg{\tens{e}}\, ,
\end{equation}
with the new resistance functions.  The corresponding particle phase Brownian stress contribution is  
\begin{equation}
  \label{eq:newsigB}
  n\avg{\widetilde{\bsig}^\mathrm{B}}^\mathrm{p} = -n\kT\tI -n \kT \langle \grad{}_c \cdot ( \widetilde{\boldsymbol{\mathfrak{R}}}_{\sigma{\cal U}} \cdot\Rfu^{-1}) \rangle\, .
\end{equation}
Combining $n\avg{\bsig^\rP}^\rp$, $n\avg{\widetilde{\bsig}^\rH}^\rp$, and $n\avg{\widetilde{\bsig}^\rB}^\rp$ defines a new particle-phase stress $n\avg{\widetilde{\bsig}}^\rp$ for colloidal suspensions.  Thus, we see that different partitions of HIs, \eg\ Eqs.~\eqref{eq:fp-nott} and \eqref{eq:fij-new}, can lead to different particle-phase stresses.  Our two particle example shows that the new partitioning is needed in order to avoid the unphysical divergence of the particle-phase stress from two colliding particles.  And we show below that this new partitioning recovers the correct thermodynamic osmotic pressure in hard-sphere suspensions.


\section{Computational Method}
\label{sec:method}

We use dynamic simulations to quantitatively investigate the particle-phase stresses $n\avg{\bsig^\rE}^\rp$, $n\avg{\widetilde{\bsig}^\rE}^\rp$, $n\avg{\bsig^\rB}^\rp$, and $n\avg{\widetilde{\bsig}^\rB}^\rp$ defined in Sec.~\ref{sec:balances} and \ref{sec:part-hydr-inter}.  For simplicity, the particles only interact through  HIs and  excluded volume effects.  For this case, \citet{brady1993a} showed that all the stresses in the system are due to the HIs, which we use the conventional Stokesian Dynamics (SD) to compute~\citep{brady-sd-ew_jfm_88,stokesian-dynamics_brady_anfm1988,bossisbrady89}.  Conventional SD evaluates the grand resistance tensor $\boldsymbol{\mathcal{R}}$ in Eq.~\eqref{eq:res-part} as
\begin{equation}
  \label{eq:sd-res}
  \boldsymbol{\mathcal{R}} = (\boldsymbol{\mathcal{M}}^\infty)^{-1} + \boldsymbol{\mathcal{R}}_\mathrm{2B} - \boldsymbol{\mathcal{R}}_\mathrm{2B}^\infty\, ,
\end{equation}
where $\boldsymbol{\mathcal{M}}^\infty$ is the far-field mobility tensor constructed  pairwisely from the multipole expansion and Fax\'{e}n's laws of the Stokes equations up to the stresslet level, and $(\boldsymbol{\mathcal{R}}_\mathrm{2B} - \boldsymbol{\mathcal{R}}_\mathrm{2B}^\infty)$ is the pairwise near-field lubrication correction with the far-field contributions removed.  The formalism of SD in Eq.~\eqref{eq:sd-res} captures two key aspects of HIs: the long-range and non-pairwise-additive nature through the inversion of $\boldsymbol{\mathcal{M}}^\infty$, and the singular behaviors when particle surfaces are close from the explicit addition of $(\boldsymbol{\mathcal{R}}_\mathrm{2B} - \boldsymbol{\mathcal{R}}_\mathrm{2B}^\infty)$.  SD recovers the exact solution of the two-particle problems and has been shown to agree well with the exact solution of three-body problems~\citep{three-spheres_wilson_jcp2013}. Moreover, our SD implementation also computes particle pressure moments in the grand resistance tensor~\citep{sd-bidisperse_wang_jcp2015}, and this allows us to quantify various pressure contributions.

We consider  dynamic simulations of monodisperse Brownian suspensions in simple shear flow without external forces or torques.  The ratio of the convective time scale $\gamd^{-1}$ and the single-particle diffusion time scale $ a^2/d_0$ defines the P\'{e}clet number $\pe = a^2\gamd/d_0$, where the imposed strain rate is $\gamd$ and the single-particle diffusivity $d_0=\kT/(6\pi\eta_0 a)$.  Integrating the particle Langevin equation, the suspension configuration changes according to  
\begin{equation}
  \label{eq:conf-change}
  \begin{split}
    \Delta \bX = & \left[\boldsymbol{\mathcal{U}}^\infty + \Rfu^{-1}\cdot\Rfe : \avg{\tens{e}} \right] \Delta t \\
    & + \kT\grad{}_c\cdot \Rfu^{-1} \Delta t +  \Delta \bX^\mathrm{B}\, ,
   \end{split}
\end{equation}
where $\Delta \bX$ is the suspension configuration change including both the positional and rotational degrees of freedom, and the stochastic Brownian displacement $\Delta \bX^\mathrm{B}$ satisfies
\begin{equation}
  \label{eq:xb-expre}
  \overline{\Delta \bX^\mathrm{B}} = 0 \text{ and }\overline{\Delta \bX^\mathrm{B} 
\Delta \bX^\mathrm{B}} = 2\kT \Delta t \Rfu^{-1}\, .
\end{equation}
The second term on the right hand side of Eq.~\eqref{eq:conf-change} is the deterministic configuration drift due to the configuration-dependent Brownian forces in Eq.~\eqref{eq:fb}.

The conventional SD method explicitly computes the grand resistance tensor $\boldsymbol{\mathcal{R}}$ with a computation cost of $\bigO(N^3)$, and is therefore only suitable for small systems.  However, the explicit construction of $\boldsymbol{\mathcal{R}}$ allows convenient calculation of the resistance pairwise hydrodynamic force $\bff^\mathrm{h}_{ij}$ in Eq.~\eqref{eq:fp-nott} and therefore $n\avg{\bsig^\rE}^\rp$ and $n\avg{\bsig^\rB}^\rp$.  The computation of the perturbative pairwise hydrodynamic force $\widetilde{\bff}^\rh_{ij}$ in Eq.~\eqref{eq:fij-new} for \emph{all} particles scales as $\bigO(N^4)$ because for each of the $N$ perturbed particle configurations, the perturbed grand resistance tensor $\boldsymbol{\mathcal{R}}'$ has to be evaluated.  This scaling further limits the system size for dynamic simulations when $n\avg{\widetilde{\bsig}^\rE}^\rp$ and $n\avg{\widetilde{\bsig}^\rB}^\rp$ are evaluated.

Our dynamic simulations largely follow~\citet{sd-brownian-susp_brady_jfm2000} with a system size $N=30$, which is adequate for rheological studies and, more importantly, for evaluating the relative merits of the two methods to partition the hydrodynamic forces.  The configuration divergence, $\grad{}_c\cdot$, is computed using the modified mid-point scheme of~\citet{asd-brownian_banchio_jcp2003}.  The particle configurations are evolved according to Eq.~\eqref{eq:conf-change}, and the time is scaled with $a^2/d_0$ when $\pe<1$ and with $\gamd^{-1}$ when $\pe\geq 1$.  Each simulation lasts $150$ dimensionless time units for $\pe=0$ and $200$ time units for $\pe> 0$ with a step size of $10^{-3}$.  When reporting the results, we discard the initial $50$ time units to ensure  steady-state behavior.

\section{Results and Discussion}
\label{sec:results}

\subsection{Equilibrium osmotic pressures}

We first study the osmotic pressure of equilibrium Brownian suspensions ($\pe=0$) at various volume fractions.  Here, the suspension osmotic pressure is the trace of the Brownian stress, \ie, $\Pi = -\tfrac{1}{3}n\avg{\tS^\rB}:\tI$.  The particle phase osmotic pressures can be similarly defined: for the resistance partition in Eq.~\eqref{eq:fp-nott}, $\Pi^\rp = -\tfrac{1}{3}n\avg{\bsig^\rB}^\rp:\tI$ with $n\avg{\bsig^\rB}^\rp$ from Eq.~\eqref{eq:sigB}, and for the perturbative partition in Eq.~\eqref{eq:fij-new}, $\widetilde{\Pi}^\mathrm{p} = -\tfrac{1}{3}n\avg{\widetilde{\bsig}^\rB}^\rp:\tI$ with $n\avg{\widetilde{\bsig}^\rB}^\rp$ from Eq.~\eqref{eq:newsigB}.

\begin{figure}
  \centering
  \includegraphics[width=3in]{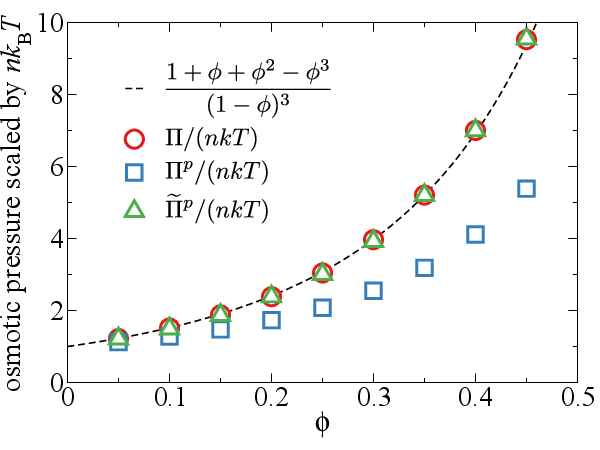}
  \caption{The suspension osmotic pressures $\Pi$, and the particle-phase osmotic pressures from the resistance and the perturbative approaches, $\Pi^\rp$ and $\widetilde{\Pi}^\rp$ respectively, as functions of the volume fraction $\phi$ for hard-sphere Brownian suspensions at $\pe = 0$.  The dashed line shows the Carnahan-Starling equation of state.}
  \label{fig:phi-pi}
\end{figure}

Fig.~\ref{fig:phi-pi} shows the dynamic simulation results for $\Pi$,  $\Pi^\mathrm{p}$, and $\widetilde{\Pi}^\mathrm{p}$ as functions of the volume fraction $\phi$ up to the fluid-solid transition density of monodisperse suspensions.  For comparison, the Carnahan-Starling equation of state for the hard-sphere fluid~\citep{colloidal_dispersions_1989} is the dashed line.  The SD suspension osmotic pressure $\Pi$ in red circles agrees excellently with the Carnahan-Starting predictions, and this agreement validates our simulation program~\citep{brady1993a, pressure-sd_morris_jor2008}.  More importantly, both particle phase osmotic pressures $\Pi^\mathrm{p}$ and $\widetilde{\Pi}^\mathrm{p}$ are more than the kinetic part $n\kT$, and $\widetilde{\Pi}^\mathrm{p}$ is almost indistinguishable from $\Pi$ over the entire $\phi$ range, \ie, $\avg{\tS^\mathrm{B}} = \avg{\widetilde{\bsig}^\mathrm{B}}^\mathrm{p}$.

The results of the particle-phase osmotic pressures $\Pi^\rp$ and $\widetilde{\Pi}^\rp$ in Fig.~\ref{fig:phi-pi} are significant.  They quantitatively show that both $\avg{\bsig^\mathrm{B}}^\mathrm{p}$ and $\avg{\widetilde{\bsig}^\mathrm{B}}^\mathrm{p}$ non-trivially contribute to the suspension Brownian stress.  This differs from the analysis of \citet{susbal_nott_physfluid2011}, where they argued that the particle phase only contributes the kinetic part, $-n\kT\tI$, to the suspension Brownian stress.  Moreover, the result $\Pi=\widetilde{\Pi}^\rp$ validates the thermodynamic arguments presented in Sec.~\ref{sec:Into} that the particles contribute the \emph{entirety} of the Brownian stress and that the fluid plays no role in the driving force for particle migration.  Therefore, to model particle migration in Brownian suspensions, the entire suspension Brownian stress can be simply added to the non-Brownian particle-phase stress.  This approach was adopted by \citet{migration-brownian-pres_morris_jfm2003} and~\citet{nmr-brownian-migration_codd_pof2009},  and we have now quantitatively shown that it is well justified.

Fig.~\ref{fig:phi-pi} also demonstrates that with non-pairwise-additive interactions, different pairwise partitions lead to different conclusions.  For example, from $\Pi^\mathrm{p}$, which is derived from the resistance partition $\bff^\mathrm{h}_{ij}$, we may conclude that the fluid phase also contributes to the suspension Brownian stress, which is inconsistent with the thermodynamics perspective.  From a purely mechanical perspective, both partitions $\bff^\mathrm{h}_{ij}$ and $\widetilde{\bff}^\mathrm{h}_{ij}$ appear reasonable.  In addition to the argument at the pair level about the divergence of the stress at contact, the distinction becomes apparent when thermal fluctuations are introduced.  From thermodynamics, which is independent of the computational details, we conclude that $\widetilde{\bff}^\mathrm{h}_{ij}$ from the perturbative approach is correct and thermodynamically consistent.  Moreover, the thermodynamic perspective is not limited to HIs and can be used to justify other ``ambiguous'' stress definitions~\citep{admal_elasticity_JElast10}.

\subsection{Rheology of sheared colloidal suspensions}

\begin{figure*}
  \centering
  \includegraphics[width=5in]{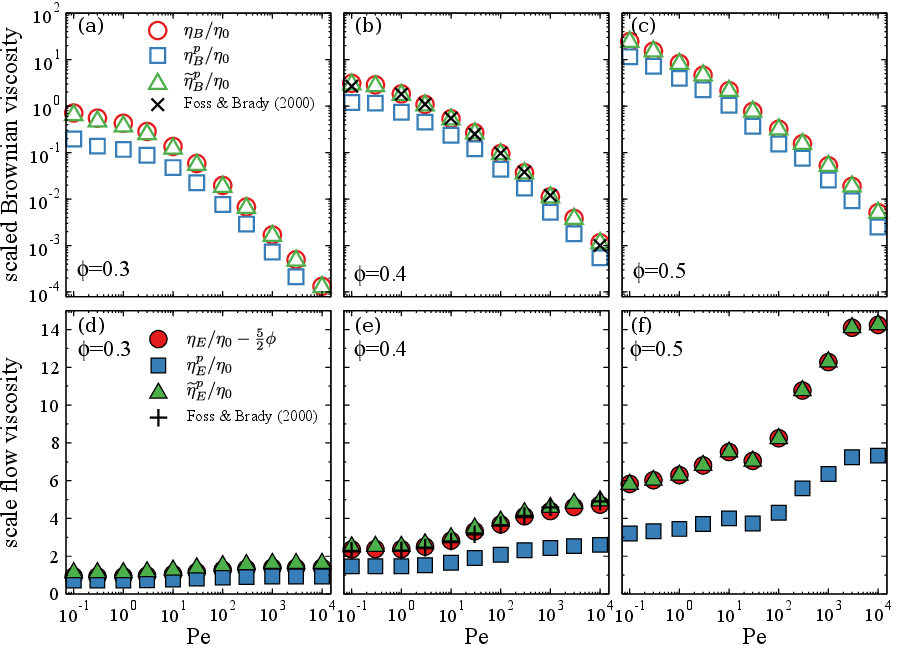}
  \caption{The suspension and particle-phase viscosities: (a)--(c): the Brownian viscosities $\eta_\rB$, ${\eta}_\mathrm{B}^\mathrm{p}$, and $\widetilde{\eta}_\mathrm{B}^\mathrm{p}$; (d)--(f): the flow viscosities $\eta_\rE-\tfrac{5}{2}\phi\eta_0$, ${\eta}_\mathrm{E}^\mathrm{p}$, and $\widetilde{\eta}_\mathrm{E}^\mathrm{p}$ as functions of the P\'{e}clet number $\pe$ at volume fractions (a,d): $\phi=0.3$, (b,e): $\phi=0.4$, and  (c,f): $\phi=0.5$.  At $\phi=0.4$ the SD results of \citet{sd-brownian-susp_brady_jfm2000} are also shown.
}
  \label{fig:eta-brownflow}
\end{figure*}

\begin{figure*}
  \centering
 \includegraphics[width=5in]{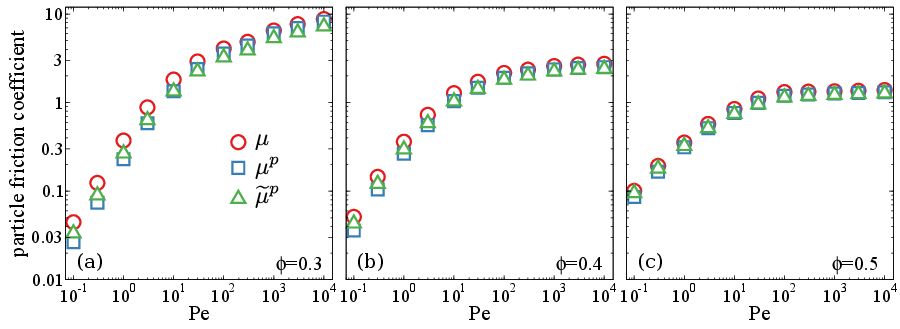}
  \caption{(The suspension friction coefficient $\mu$ and the particle-phase friction coefficients $\mu^\mathrm{p}$ and $\widetilde{\mu}^\mathrm{p}$, as functions of P\'{e}clet number, Pe, at volume fractions (a): $\phi=0.3$, (b): $\phi=0.4$, and (c): $\phi=0.5$.  Here, the friction coefficients are defined without the solvent contribution.}
  \label{fig:mup}
\end{figure*}

Next, we turn to the suspension dynamics for $\pe > 0$.  In Fig.~\ref{fig:eta-brownflow} and \ref{fig:mup} we present different aspects of suspension rheology from the suspension stress $\bSig^\mathrm{(p)}$ and the particle-phase stresses $n\avg{\bsig^\mathrm{h}}^\mathrm{p}$ and $n\avg{\widetilde{\bsig}^\mathrm{h}}^\mathrm{p}$ over a wide range of $\pe$ for $\phi = 0.3$, $0.4$, and $0.5$.

For hard-sphere Brownian suspensions, the viscosity from the ``particle stress'' $\bSig^\mathrm{(p)}$ can be split according to Eq.~\eqref{eq:susp-stress} into a Brownian part, $\eta_{\mathrm{B}}=n\avg{S^\mathrm{B}}_{12}/\gamd$, and a flow part, $\eta_{\mathrm{E}}=n\avg{S^\mathrm{E}}_{12}/\gamd$, where the subscript ``$12$'' represents the flow-flow gradient component of the stress.  With $n\avg{\bsig^\rE}^\rp$ and $n\avg{\bsig^\rB}^\rp$ in Eqs.~\eqref{eq:stresslet-E} and \eqref{eq:sigB}, respectively, we can similarly define the particle-phase viscosities $\eta_{\mathrm{B}}^\mathrm{p}$ and $\eta_{\mathrm{E}}^\mathrm{p}$, and for the perturbative particle-phase stresses, $\widetilde{\eta}_{\mathrm{B}}^\mathrm{p}$ and $\widetilde{\eta}_{\mathrm{E}}^\mathrm{p}$, from Eqs.~(\ref{eq:newsigE}) and (\ref{eq:newsigB}).  Among these viscosities, only the suspension $\eta_{\mathrm{B}}$ and $\eta_{\mathrm{E}}$ are experimentally accessible.  Moreover, since the direct hydrodynamic particle-phase stresses, $n\avg{\bsig^\rE}^\rp$ and $n\avg{\widetilde{\bsig}^\rE}^\rp$, do not contain contributions from an isolated particle, to compare the particle phase viscosities $\eta^\rp_\rE$ and $\widetilde{\eta}^\rp_\rE$ with the suspension viscosity $\eta_\rE$, it is necessary to remove the single-particle (Einstein) viscosity contribution $\tfrac{5}{2}\phi\eta_0$ from $\eta_\rE$.

Fig.~\ref{fig:eta-brownflow} shows various Brownian and flow viscosities as functions of $\pe$.  Also presented are the results of~\citet{sd-brownian-susp_brady_jfm2000} at $\phi = 0.4$.  For the Brownian viscosities in Fig.~\ref{fig:eta-brownflow}a--\ref{fig:eta-brownflow}c,  $\eta_\mathrm{B}$,  $\eta_\mathrm{B}^\mathrm{p}$, and $\widetilde{\eta}_\mathrm{B}^\mathrm{p}$ exhibit shear-thinning with increasing $\pe$, and at the same $\pe$, they increase significantly with $\phi$.  Consistent with Fig.~\ref{fig:phi-pi}, the perturbative particle-phase Brownian viscosity $\widetilde{\eta}_\mathrm{B}^\mathrm{p}$ agrees excellently with the suspension Brownian viscosity $\eta_\mathrm{B}$ for all $\pe$ and $\phi$, while the resistance particle-phase Brownian viscosity ${\eta}_\mathrm{B}^\mathrm{p}$ is significantly lower.  

The suspension flow viscosity $\eta_\rE$ and the particle-phase flow viscosities  $\eta_\rE^\rp$ and $\widetilde{\eta}_\rE^\rp$ in Fig.~\ref{fig:eta-brownflow}d--\ref{fig:eta-brownflow}f show the expected shear-thickening behaviors; that is, these viscosities grow with increasing $\pe$.  The shear-thickening is most pronounced at $\phi = 0.5$, and the dip at $\pe=30$ in Fig.~\ref{fig:eta-brownflow}f is most likely due to suspension structural changes associated with the small system size $N=30$.  These figures show that the perturbative particle-phase viscosity $\widetilde{\eta}_\rE^\rp$ closely follows the suspension flow viscosity $\eta_\rE$ with the particle self-contribution, $\tfrac{5}{2}\phi\eta_0$, removed across different volume fractions, \ie, $\widetilde{\eta}_\rE^\rp \approx \eta_\rE-\tfrac{5}{2}\phi\eta_0$, and the agreement improves with increasing $\phi$.  

As $\pe\rightarrow 0$, the flow viscosity corresponds to the high-frequency dynamic shear viscosity of an equilibrium suspension.  That $\eta_\mathrm{E}$ and $\widetilde{\eta}_\rE^\rp$ differ in this limit is expected as $\eta_\mathrm{E}$ contains the single-particle Einstein viscosity; without this term the two are indistinguishable.   Away from this limit, the suspension viscosity $\eta_\rE$ and the perturbative particle-phase viscosity $\widetilde{\eta}^\rp_\rE$ respond to the suspension structural distortion due to the imposed flow almost identically, \ie, $[\widetilde{\eta}^\rp_\rE(\pe) - \widetilde{\eta}^\rp_\rE(\pe\rightarrow0)]$ is the same as $[\eta_\rE(\pe) - \eta_\rE(\pe\rightarrow0)]$ at all $\phi$.  Clearly, the perturbative particle-phase stress $n\avg{\widetilde{\bsig}^\rE}^\rp$ responds to the microstructural changes as $n\avg{\tS^{\rE}}$ does.  On the other hand, the resistance particle-phase viscosity $\eta_\rE^\rp$ is much lower than $\eta_\rE$ and $\widetilde{\eta}_\rE^\rp$, and exhibits less shear thickening with increasing $\pe$.  

The ratio of the suspension shear stress $\Sigma^\mathrm{(p)}_{12}$ to the osmotic pressure $\Pi$ defines the suspension friction coefficient $\mu = \Sigma^\mathrm{(p)}_{12}/\Pi$, which is a macroscopic material property traditionally used to characterize the rheology of granular materials~\citep{review-granular-flow_forterre_arfm2008}, and more recently, suspensions~\citep{boyer-granular-rheology_prl2011, confine-friction-rheo_garagash_jfm2014, Wang_yield-glass_2015}.  The particle-phase friction coefficients can be similarly defined from $n\avg{\bsig^\rh}^\rp$ as $\mu^\rp = n\avg{\sigma^\rh}^\rp_{12}/\Pi^\rp$ and from $n\avg{\widetilde{\bsig}^\rh}^\rp$ as $\widetilde{\mu}^\rp = n\avg{\widetilde{\sigma}^\rh}^\rp_{12}/\widetilde{\Pi}^\rp$.  These friction coefficients characterize the overall flow behaviors of the suspension.

Fig.~\ref{fig:mup} presents the suspension friction coefficient $\mu$ and the particle-phase friction coefficients $\mu^\rp$ and $\widetilde{\mu}^\rp$ as functions of P\'{e}clet number at volume fractions $\phi=0.3$, $0.4$, and $0.5$.  For $\pe\ll 1$,  $\mu$ is proportional to $\pe$ as the normal stress is dominated by the equilibrium osmotic pressure but the shear stress increases with $\pe$.  For $\pe\gg 1$, $\mu$ asymptotes to a constant, because in this limit both the shear and the normal stresses are proportional to $\pe$.  Comparing Fig.~\ref{fig:mup}a--\ref{fig:mup}c at different $\phi$ reveals that the high $\pe$ limiting friction coefficient decreases with $\phi$, consistent with experiments of \citet{boyer-granular-rheology_prl2011} on non-Brownian suspensions.  On the other hand, when $\pe\ll 1$, the $\mu$-$\pe$ slope increases with $\phi$ due to the significant increase in suspension viscosity at higher $\phi$.

The particle-phase friction coefficients $\mu^\rp$ and $\widetilde{\mu}^\rp$ agree well with each other and with the suspension friction coefficient $\mu$.  Moreover, their agreement improves with increasing $\phi$.  Since friction coefficients are ratios of different stresses components, the agreement in Fig.~\ref{fig:mup} shows that these stresses respond consistently to the imposed flow, and demonstrates another benefit of rheological characterization via friction coefficients---different stress definitions indeed give the same suspension response.

\section{Concluding Remarks}
\label{sec:conclusions}

The computational results in Sec.~\ref{sec:results} show that the flow contribution to the perturbative particle-phase stress, $n\avg{\widetilde{\bsig}^\rE}^\rp$, is almost indistinguishable from the suspension stresslet $n\avg{\tS^\rE}$ without the single-particle contribution, and the Brownian contribution satisfies $n\avg{\widetilde{\bsig}^\rB}^\rp =  n\avg{\tS^\rB}$ in both equilibrium and non-equilibrium sheared suspensions.  Clearly, the suspension stress $\bSig^\mathrm{(p)}$ without the single-particle hydrodynamic contribution satisfactorily approximates the experimentally inaccessible particle phase stress $n\avg{\widetilde{\bsig}^\rh}^\rp$.  That $\bSig^\mathrm{(p)}\approx n\avg{\widetilde{\bsig}^\rh}^\rp$ explains the success of existing SBM in many applications.  The approximate equivalence between $\bSig^\mathrm{(p)}$ and $n\avg{\widetilde{\bsig}^\rh}^\rp$ also revives the physical soundness of the SBM for the shear-induced particle migration and other studies.

Our conclusions critically rely on the correct partition of the HIs when defining the particle-phase stress.  
We have shown that separate physical arguments, such as thermodynamics for Brownian systems and the nondivergence of the stress at the pair level, are necessary for choosing the correct partition.  In this work, the configurational perturbation partition in Eq.~\eqref{eq:fij-new} is thermodynamically consistent and is the key to the revival of the SBM.

The near perfect agreement between the suspension and particle-phase stresses can be anticipated from the stresslet definitions in Eqs.~(\ref{eq:stresslet}) and (\ref{eq:stresslet-E}) (or Eq.~(\ref{eq:newsigE})).  The suspension stress requires $\Rsu$, while the particle-phase needs $\boldsymbol{\mathfrak{R}}^{\sigma\mathrm{U}}_{ij}=\tfrac{1}{2}\br_{ij} \uRfu_{ij}$.  When two particles are near touching the strong lubrication forces are localized at the point of contact and $\Rsu =\tfrac{1}{2}\br \uRfu$ as $\br \rightarrow 2a$, as shown by \citet{brady1993a}.  For dense suspensions when lubrication interactions dominate, the suspension and particle-phase stresses must be the same.  Evidently, the configuration perturbation partitioning of the hydrodynamic forces captures the correct long-range, many-body interactions resulting in near identity between the suspension and particle-phase stresses.   Thus, one may with confidence use the {\em experimentally} measurable suspension stress in the particle-phase momentum balance and the resulting particle-phase flux as long as the isolated single-particle Einstein viscosity contribution is removed from the suspension stress.  This is the {\em only} change needed.

We should note that the excellent agreement between the suspension and properly partitioned particle-phase stress has been established for relatively concentrated suspensions ($\phi \ge 0.1$).  For very dilute suspensions, detailed pair-level HIs may be important for predicting particle motion.  This applies equally to the drag coefficient $\bzeta$ and the particle-stress in Eq.~\ref{eq:jrel}.

We have resolved several issues in the analysis of \citet{susbal_nott_physfluid2011} including the agreement between the SBM and experiment and the role of Brownian motion in the particle-phase stress.  From our results, we believe the SBM is, and will remain, a physical and valid approach to investigate the behavior of flowing suspensions.  SBM-based constitutive modeling of suspensions  can  continue with confidence.

\bigskip


\begin{acknowledgements}
M.W. gratefully acknowledges support from the Natural Sciences and Engineering Research Council of Canada (NSERC) by a Postgraduate Scholarship (PGS), and the National Science Foundation (NSF) grant CBET-1337097. T.Z. gratefully acknowledges support from the Department of Energy (DOE) grant DE SCO22966.
\end{acknowledgements}

\bibliographystyle{apsrev4-2}
\bibliography{ref}

\end{document}